\documentclass[12pt,prb,aps]{revtex4}

\usepackage{amsmath}
\usepackage{amssymb}
\usepackage{amsfonts}
\usepackage{graphicx}

\newcommand{\cit}{California Institute of Technology, Department of Physics,
Pasadena, CA, 91125}

\begin{document}

\preprint{}

\title{Macroscopic coherence effects in a mesoscopic system:  Weak localization of thin silver films in an undergraduate lab}
\author{A. D. Beyer}
 \affiliation{\cit}
 \email{beyer@its.caltech.edu}
 \author{M. Koesters}
  \affiliation{\cit}
\author{K.G. Libbrecht}
 \affiliation{\cit}
 \author{E.D. Black}
 \affiliation{\cit}

\begin{abstract}
We present an undergraduate lab that investigates weak localization
in thin silver films.  The films prepared in our lab have thickness,
$a$, between $60-200$~\AA, a mesoscopic length scale. At low
temperatures, the inelastic dephasing length for electrons,
$L_{\phi}$, exceeds the thickness of the film ($L_{\phi} \gg a$),
and the films are then quasi-2D in nature. In this situation, theory
predicts specific corrections to the Drude conductivity due to
coherent interference between conducting electrons' wavefunctions, a
macroscopically observable effect known as weak localization. This
correction can be destroyed with the application of a magnetic
field, and the resulting magnetoresistance curve provides
information about electron transport in the film. This lab is
suitable for Junior or Senior level students in an advanced
undergraduate lab course.
\end{abstract}

\maketitle

\section{Introduction}

One of the many challenges in a senior level lab course is to
simulate the research environment.  The labs generally should be
geared toward helping students transition from performing
pre-packaged experiments to more independent experiments. However,
experiments of this sort should be designed so that motivated
students can succeed and can do so in a term or semester. With this
in mind, we have designed a simple experiment for students to
observe the phenomena of weak localization.

Weak localization is a macroscopically observable consequence of the
quantum mechanical behavior of electrons. Electrons begin to
localize around impurities at low temperatures because of an effect
that can be described as a self-interference effect. As electrons
scatter around during transport measurements, they hit impurities,
and special paths for partially scattered electron waves add
together to start localization of electrons. This effect of
pre-localization is lumped under the term weak localization, and one
consequence is that the bulk behavior of electrons is altered.

The bulk property that most clearly gives evidence for weak
localization is the magnetoconductance signal, which is best seen in
thin film samples at low temperatures (accessible with liquid
helium). The magnetoconductance is a small correction to the bulk
conductance of a sample, theoretically on the order of $10^{-4}$ of
the bulk conductance. In this lab, we discuss how samples are made
and how the magnetoconductance is measured.

One of the goals of this lab is to teach students techniques for
measuring small signals.  Measuring small signals is a common
occurrence in condensed matter experiments, and for experimental
physicists in general, it is good to learn tricks for working with
faint signals. In our experiment, the magnetoresistive signal we
want to measure is swamped by the bulk resistance of the sample. To
magnify this weak signal, we use a resistance bridge. The philosophy
is simple: null out the large resistance signal due to the bulk
conductance and amplify the remaining signal due to the
magnetoresistance. In addition to discussing the bridge, we also
review sources of noise that can arise in the bridge, such as ground
loops, and how to design the bridge to eliminate them.

Typical students' results and a brief interpretation of these
results are then discussed.  We find that for the Ag thin film
samples made in our lab, students observe weak localization with
spin effects at temperatures below about 14 K. Above 14 K, phonons
destroy the spin effects, and spinless weak localization is
observed.

\section{Theory}

Weak localization is a correction to the Drude conductivity.  Before
we discuss the correction, let us review the basic Drude prediction
and the physical assumptions that lead to it.

\subsection{Drude conductivity}

If we apply an electric field inside a normal metal, that field will
drive an electrical current. The current density $\mathbf{J}$ is
related to the field $\mathbf{E}$ by
\[
\mathbf{J} = \sigma \mathbf{E}
\]
where $\sigma$ is called the conductivity. Materials with high
conductivity allow large currents for a given field, whereas
materials with lower conductivities allow lower current densities.

A formula for the conductivity was derived by
Drude~\cite{ref:Drude}, based on some very simple assumptions about
the microscopic properties of metals. If the current is carried by
electrons, then the current density is just
\[
\mathbf{J} = -n e <\mathbf{v}>
\]
where $n$ is the number of electrons per unit volume, $-e$ is the
electric charge on a single electron, and $<\mathbf{v}>$ is the
average velocity of the electrons. Now, we need to calculate this
average velocity. To do so, let's assume that the electrons move
inside the metal ballistically until they run into something. The
electric field provides a force on the electrons given by
$-e\mathbf{E}$, and the equation of motion is then
\[
m \frac{d \mathbf{v}}{dt} = - e \mathbf{E}
\]
If we know an electron's velocity at time $t=0$, we can calculate it
at a later time $t$, provided it does not run into anything.
\[
\mathbf{v}(t) = \mathbf{v}(0) - \frac{e \mathbf{E} t}{m}
\]
For this particular electron, let's assume that just before $t=0$ it
ran into something inside the metal (an impurity or phonon, for
example) and bounced off. Moreover, let's assume that, just after
the collision, the velocity of the electron is completely random.
This means that, on average,
\[
<\mathbf{v}(0)> = 0
\]
With the assumption of isotropic scattering, the average velocity of
all the electrons inside a metal is easy to calculate.
\[
<\mathbf{v}> = - \frac{e \mathbf{E} \tau}{m}
\]
where $\tau$ is the average time since the last scattering event.
This ensemble average is independent of the time when we take the
snapshot, and it gives the conductivity.
\[
\sigma_{D} = \frac{n e^2 \tau}{m}
\]

This expression is known as the Drude conductivity, after the
scientist who first derived it. It relies on the assumption that,
after a scattering event, an electron's velocity is completely
randomized. Put another way, the electron has no memory of its
previous state after it suffers a collision inside the material. As
we shall see, this assumption is good at room temperatures, but it
breaks down when the temperature of the metal becomes very low. The
cause of this breakdown is the wave nature of the electron, which
did not play a role in the derivation of the Drude conductivity, and
the resulting change in the conductivity is one of the few
macroscopically observable consequences of the wave-particle duality
of matter.

\subsection{Coherent backscattering}

\begin{figure}
\begin{center}
\includegraphics{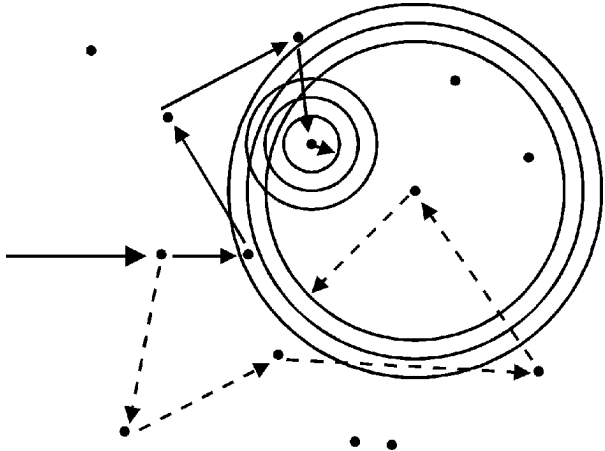}
\caption{\label{fig:BeyerFig1} Most randomly selected partial waves
have random relative phases, and on average they add incoherently.
In this figure, two scattering paths are considered---one shown by a
solid line and one by a dashed line.   Notice that the wavefronts
for both scattering paths are all separated by the same radial
length. This separation is the Fermi wavelength, $\lambda_{F}$. For
these randomly scattered partial waves, the wavefronts do not align,
so on average there will be no net interference.}
\end{center}
\end{figure}

\begin{figure}
\begin{center}
\includegraphics{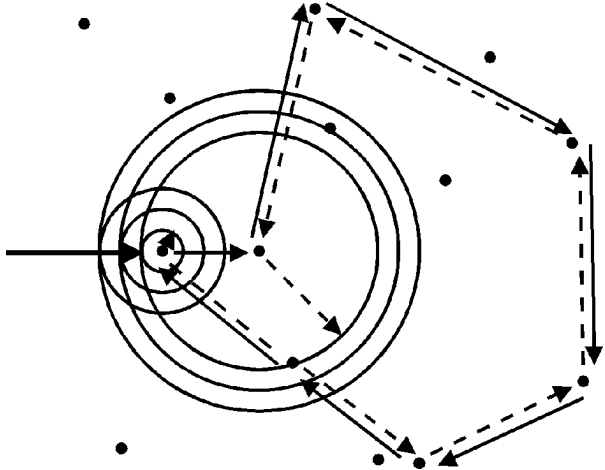}
\caption{\label{fig:BeyerFig2} Each partial wave that returns to the
origin has another partial wave that is in phase with it, in the
backscatter direction. Again, we consider two partially scattered
waves. Unlike Figure \ref{fig:BeyerFig1}, the two paths are the same
except that they are traversed in opposite order (i.e. if we apply
time reversal to one path, we get the second path). This similarity
and the fact that their net scattering direction is in the
backscattering direction causes the wavefronts of the two paths to
align on average, and there is a net interference effect. This means
on average the backscatter probability is modified from the
predictions of classical physics.}
\end{center}
\end{figure}

When we calculated the Drude conductivity, we assumed the electrons
were point particles, obeying the laws of classical physics. In
reality, however, electrons have associated with them wave
functions, the square of the amplitude of which gives the
probability of observing an electron. When the wave that describes
an electron scatters off of some obstacle in its path, it produces
partial waves emanating from that obstacle, much like ripples in a
pond produced by a wave when it hits a stationary reed sticking out
of the surface. These partial waves go on to strike other obstacles,
impurities or defects in the case of a metal, and produce more
partial waves. All of these partial waves add up, as waves do, to
produce a complicated diffraction or interference pattern. For the
most part, the phase between any two partial waves is random, and
the partial waves, on average, add incoherently (see Figure
\ref{fig:BeyerFig1}). However, there is one direction in which the
partial waves will always add up in phase, and that is the direction
opposite that of the initial wave. This is because, for every path
that takes a partial wave back to its origin, there is a
complementary path with the same length that takes the same route,
but in the opposite direction. The two partial waves that take these
complementary paths will always add coherently (see Figure
\ref{fig:BeyerFig2}). The sum of all these complementary waves then
gives a slightly stronger wave going backward relative to the
initial direction.  This corresponds to an enhanced probability of
backscattering, which in turn reduces the conductivity below the
Drude prediction.

If the metal is infinitely large and the electrons can maintain
phase coherence over infinitely long complimentary paths, then the
backscattering effect dominates the dynamics of the electrons, and
the conductivity is completely suppressed. In this case, the metal
becomes an insulator, and the electrons are trapped, or localized,
by the disordered scattering centers. For samples with finite size,
or more commonly finite coherence lengths, the backscattering gives
rise to a small correction to the Drude conductivity. This small
correction ought to be called ``pre-localization,'' but instead it
is commonly referred to as ``weak localization.''

The coherent backscattering effect can only occur if the phase of
each partial wave is preserved as it goes around its path. At high
temperatures, where most scattering events are off of phonons,
coherent backscattering cannot occur. A magnetic field can also
introduce a phase difference between the complementary paths,
destroying the coherent backscattering and any correction to the
Drude conductivity it produces. The resulting dependence of the
conductivity on temperature or applied magnetic field is quite small
even at liquid-helium temperatures, but it can be observed
experimentally by employing a few basic low-noise techniques. In the
experiment we will describe below, we use this weak-localization
effect to teach some basic low-noise and small-signal-detection
methods that are commonly used in many research labs.

\subsection{Weak localization}

A complete, quantitative derivation of the weak localization effect
requires the use of quantum many-body theory and the
quantum-field-theory techniques that go with it. This is beyond the
scope of this paper, and we will just give the result here without
derivation. See Ref. 2 or Ref.3 for a detailed treatment of the
theory of weak localization.

\begin{figure}
\begin{center}
\includegraphics{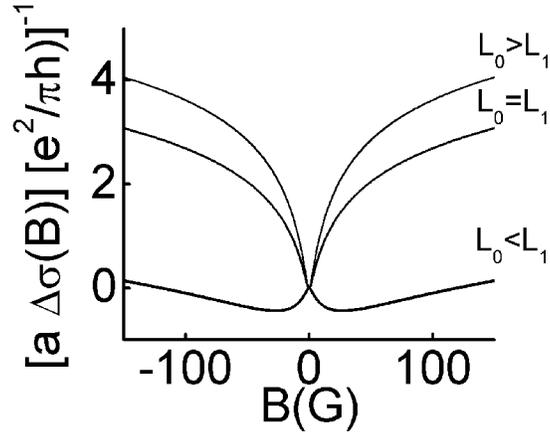}
\caption{\label{fig:BeyerFig3} The weak-localization
magnetoconductance at a fixed temperature. The different curves
represent different relative contributions from spin effects.}
\end{center}
\end{figure}

The change in the conductance of a sample when a magnetic field is
applied is called the magnetoconductance. This magnetoconductance is
easiest to observe in a two-dimensional sample, where we can apply a
magnetic field that is perpendicular to the sample and thus
perpendicular to all of the complimentary, closed-loop paths that
give rise to the coherent backscattering. For a thin film (a nearly
two-dimensional sample), the magnetoconductance is
\begin{eqnarray}
\label{eq:weak-localization}
a  \Delta\sigma(B) & = &  \frac{e^2}{\pi h} \left\{ \frac{3}{2} \left[ \psi \left( \frac{1}{2} + \frac{\hbar c}{4 e} \frac{1}{L_1^2 B} \right) - \ln \left( \frac{\hbar c}{4 e} \frac{1}{L_1^2 B} \right) \right] \right. \nonumber \\
& - & \left. \frac{1}{2} \left[ \psi \left( \frac{1}{2} +
\frac{\hbar c}{4 e} \frac{1}{L_0^2 B} \right) - \ln \left(
\frac{\hbar c}{4 e} \frac{1}{L_0^2 B} \right) \right] \right\}
\end{eqnarray}
where $a$ is the thickness of the film,
$\Delta\sigma(B)\equiv\sigma(B)-\sigma (0)$, and $\psi$ is the
digamma function, defined in terms of the ordinary gamma function
$\Gamma (x)$ as
\begin{eqnarray*}
\psi (x) & = & \frac{d}{dx} \ln \Gamma (x) \\
& = & \frac{1}{\Gamma (x)} \frac{d \Gamma (x)}{dx}
\end{eqnarray*}
A plot of Equation (\ref{eq:weak-localization}) is shown in Figure
\ref{fig:BeyerFig3}.  The curves in the figure are plotted at fixed
temperature, but they have varying contributions of spin effects,
which is accomplished by varying the ratio of $L_0$ and $L_1$.

The dephasing lengths $L_0$ and $L_1$ are combinations of the
average lengths an electron can diffuse before colliding with a
phonon $L_\phi$, and the average lengths an electron can diffuse
before it gets dephased by spin-orbit $L_{so}$ or spin-flip $L_{sf}$
interactions with the scatterers.
\[
\frac{1}{L_0^2} = \left( \frac{1}{L_\phi^2} + \frac{2}{3 L_{sf}^2}
\right) + \frac{4}{3 L_{sf}^2}
\]
and
\[
\frac{1}{L_1^2} = \left( \frac{1}{L_\phi^2} + \frac{2}{3 L_{sf}^2}
\right) + \frac{4}{3 L_{so}^2}
\]
A film is considered thin, or quasi-two-dimensional, if it is much
thinner than the typical dephasing lengths, $a \ll L_0, L_1$.

\begin{figure}
\begin{center}
\includegraphics{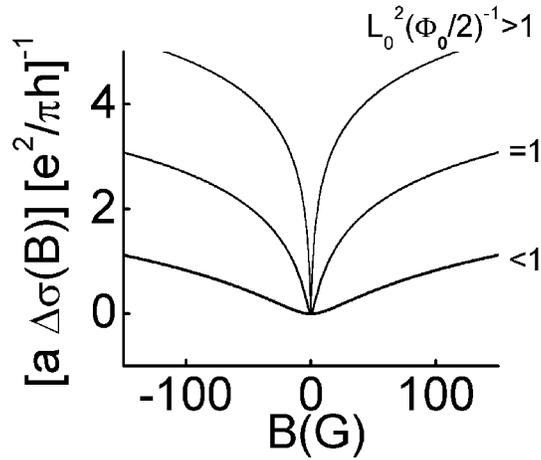}
\caption{\label{fig:BeyerFig4} The weak-localization
magnetoconductance in a sample where spin effects are negligible, at
several different temperatures ($\Phi_{0}=\frac{\hbar c}{4e}$).}
\end{center}
\end{figure}

If spin effects are negligible $L_0 = L_1 = L_\phi$, and the weak
localization magnetoconductance is particularly simple.
\begin{equation}
\label{eq:spinless-weak-localization} a  \Delta\sigma(B) =
\frac{e^2}{\pi h} \left\{  \left[ \psi \left( \frac{1}{2} +
\frac{\hbar c}{4 e} \frac{1}{L_\phi^2 B} \right) - \ln \left(
\frac{\hbar c}{4 e} \frac{1}{L_\phi^2 B} \right) \right] \right\}
\end{equation}
A plot of Equation (\ref{eq:spinless-weak-localization}) at various
temperatures is shown in Figure \ref{fig:BeyerFig4}. In the figure,
temperature is varied by varying $L_0$, because the average
inelastic dephasing length depends on temperature.

For most conductors, the spin-flip interaction is negligible
($L_{sf} \gg L_{so},L_{\phi}$), and the terms proportional to
$1/L_{sf}^{2}$ can be safely neglected.  The spin-orbit term,
however, scales with the atomic number of the metal,
$1/L_{so}^{2}\sim Z^{4}$.  For light metals, such as lithium or
magnesium, spin-orbit effects are negligible.  For heavier metals,
such as silver or gold, spin-orbit effects can dramatically alter
the weak localization magnetoconductance signature.

Compared to the Drude conductivity, the weak-localization correction
is very small. At most, the fractional correction to the
conductivity, $\delta \sigma\equiv\sigma(0)-\sigma_{D}$, is of the
order
\[
 \left| \frac{\delta \sigma}{\sigma}
\right| \leq \frac{e^2}{a \pi h} \frac{m}{n e^2 \tau}
\]

For a typical metal film with a thickness of $a = 100$\AA, this is,
approximately
\[
 \left| \frac{\delta \sigma}{\sigma} \right|
\leq 10^{-4}
\]

In order to measure this magnetoconductance with an accuracy of $1
\% $, then, we would need to resolve changes in the total
conductivity on the order of parts-per-million. This resolution
demands good low-noise and small-signal detection methods, and
teaching these methods is the purpose of this lab.

\section{Experimental apparatus}

\subsection{Samples and their preparation}

The measurement electronics form the heart of this lab. Here,
students learn techniques widely used in condensed matter labs for
measuring very small signals in the presence of noise.  In
particular, they learn how to measure part-per-million changes in
the resistance of a sample that will accept no more than a few
milliamps of excitation current.

The first thing that our students do is make their thin film Ag
samples.  To make this easy for them, we have a rugged, home-built
evaporative deposition (ED) system.

The technique of ED is simple enough so that most labs are in
possession of the equipment needed to set up an ED system.   The
basic idea is to make a tightly sealed chamber and evacuate it,
using vacuum pumps, so that the pressure is low enough for air
molecules to be in the molecular flow regime (i.e. their mean free
path is longer than the dimensions of the vacuum chamber).  Under
these conditions, ED can be performed by melting a Ag pellet. The
resulting gas of Ag atoms will then travel to the walls of the
chamber with a low probability of hitting air molecules.  If a
substrate is placed on one of the walls, it will then be possible to
deposit a thin film of Ag onto this substrate. To meet the criteria
of making the film quasi-2D (i.e. $a\ll L_{\phi}$), we find films
need to be between 60-200\AA ~thick. To make sure samples are that
thin, our students use a commercially available thickness
gauge\cite{Infinicon} to monitor the thickness of their films.  So
by using ED, thin film Ag samples can easily be made.

Once a sample is made, the students have to make contact to the
film, cool it down, and make measurements in an applied magnetic
field. In our lab, we use a commercially available liquid $^{4}$He
dewar from Quantum Design\cite{qdusaDewar}, which comes with a 5
Tesla superconducting magnet, to cool the sample and make
measurements in a magnetic field.  This is not the only option
though.  The experiment can be done just as easily with an
inexpensive dipping probe, inserted directly into a liquid $^{4}$He
storage dewar.  In fact, such a system would probably be preferable.
The computer-controlled Quantum Design cryostat can sometimes
obscure the physics of the experiment, and the central focus of the
lab should be on the physics of electron transport in mesoscopic
systems and techniques for low-level signal detection.

To prepare the samples for measurements, the students need to apply
contacts to the film, and they need to have electrical connections
to the film when it is at liquid $^{4}$He temperatures (T$\geq2K$).
For our setup, students make contact from the sample to a
resistivity sample stage, called a puck\cite{qdusaDC}, available
from Quantum Design.  The contact is made using silver paste and
gold wires.  Once the contacts are made, the sample stage is loaded
into the cold dewar, and we then provide a breakout box that allows
connections to be made to the resistivity puck from inside the
dewar.

At this point, the samples are ready to be cooled, and we need a way
to measure the small magnetoconductance signal.  To do this we use a
resistive bridge technique, which we describe next.

\subsection{Detection of magnetoresistance}

The measurement electronics form the heart of this lab. Here,
students learn techniques widely used in condensed matter labs for
measuring very small signals in the presence of noise.  In
particular, they learn how to measure part-per-million changes in
the resistance of a sample that will accept no more than a few
milliamps of excitation current.

To relate the observable resistance change to the theoretical
prediction for the change in conductivity, start with the relation
between resistance and conductivity for a film of thickness $a$,
length $l$, and width $w$.
\begin{equation}
R=\frac{1}{\sigma}\frac{l}{a w} \notag\\
\end{equation}

The change in resistance due to a change in conductivity is just
\begin{equation}
\Delta R=-\frac{\Delta\sigma}{\sigma^{2}}\frac{l}{aw} \notag\\
\end{equation}
\begin{equation}
=-\frac{a\Delta\sigma}{a\sigma}R \notag\\
\end{equation}
\begin{equation}
\label{eq:magnetoresistance} =-a\Delta\sigma R_{\Box} R
\end{equation}
where we have defined $1/a\sigma\equiv R_{\Box}$ as the resistance
of a square film.  Equations (\ref{eq:weak-localization}) and
(\ref{eq:spinless-weak-localization}) give predictions for
$a\Delta\sigma\left(B\right)$, which we will compare with
measurements of $-\Delta R/ R_{\Box}R$.

To get $R$ and $R_{\Box}$, we simply pass a small excitation current
through the sample and then measure the resulting voltage across the
sample. This gives us the total resistance of the sample $R(0)$, and
from that we can calculate $R_{\Box} = R(0)\cdot \frac{w}{l}$,
provided we know the sample's length and $l$ and width $w$.

    Measuring the magnetoresistance $\Delta R(H)$ is done in a
similar way, but because $\Delta R(H)$ is so much smaller than the
``background'' signal $R(0)$, we need an extra trick: nulling.  What
we are really measuring, remember, is the voltage across the sample
produced by the excitation current $i$, or $V(H)=i R(H)$.

    If we can produce a reference voltage that is equal to the
zero-field voltage across the sample, $V_{0}=i R(0)$, then we can
subtract this from the actual voltage across the sample to get
$\Delta R(H)$.

\begin{equation}
V(H)-V_{0} = i R(H) - i R(0) \notag\\
\end{equation}
\begin{equation}
= i \Delta R(H)
\end{equation}
This very small signal can then be amplified and examined in detail
as a function of magnetic field H.

    This process of nulling will only work if we can produce a small
voltage $V_{0}$ that is independent of the magnetic field.
Fortunately, this is easy to do using a passive, adjustable, low
noise voltage divider.  We used a Dekatran DT72A tunable voltage
divider,\cite{Deka} driven by the same voltage that produced the
excitation current for the sample.

\begin{figure}
\begin{center}
\scalebox{1.0}{\includegraphics{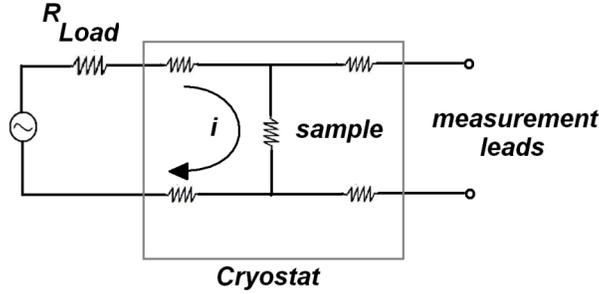}}
\caption{\label{fig:BeyerFig5}A four-wire arrangement.  Here no
current flows through the contact resistances, so when we measure a
voltage, it is due to the sample only.}
\end{center}
\end{figure}

    In both the $R_{\Box}$ and $\Delta R(H)$ measurements, it is
important not to use the same set of wires to carry the excitation
current and to measure the resulting voltage across the sample. Long
wires leading down into the cryostat may have resistances of their
own which, if their leads carry current, will produce voltages that
have nothing to do with the sample.  In order to measure only that
voltage produced by the sample, we employ a four-wire geometry, as
shown in Figure \ref{fig:BeyerFig5}.

\begin{figure}
\begin{center}
\scalebox{1.0}{\includegraphics{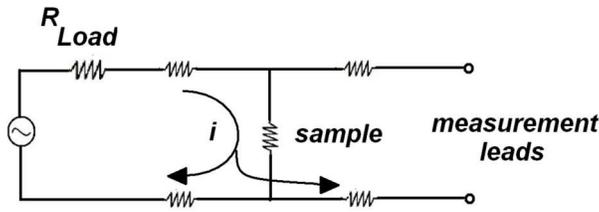}}
\caption{\label{fig:BeyerFig6}An example of how a ground loop may
develop in a four wire measurement.  The lock-in inputs we use are
floated to prevent this condition.}
\end{center}
\end{figure}

    For this four-wire technique to be effective, no current must be
allowed to flow along the measurement leads.  If it did, a spurious
voltage would be produced from the resistances in the measurement
leads, and this would contaminate the measurement.  Such a condition
can result, for example, from the measurement device and the
excitation source sharing a common ground, as shown in Figure
\ref{fig:BeyerFig6}. This condition is known as a ground loop, and
care is needed to avoid it.

    All of these measurements are performed at audio frequencies
using a lock-in, which the students have been introduced to in a
previous lab.\cite{Ph77LockIn} We add to their training with a
lock-in by encouraging them to analyze the noise in each component
of their apparatus and in the apparatus as a whole.  This is easily
done with a lock-in by terminating the input of a device with a
$50\Omega$ terminator and then measuring the noise of the output
signal on the lock-in.  The students can then determine which
component is setting their noise floor;  our students find the
pre-amplifier sets their noise floor.

\begin{figure}
\begin{center}
\scalebox{1.0}{\includegraphics{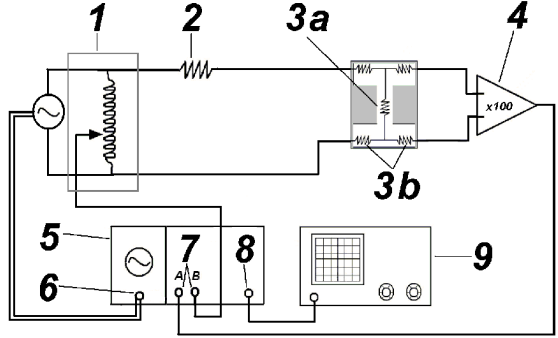}}
\caption{\label{fig:BeyerFig7}The setup used to measure weak
localization.  It is a 4-wire resistance bridge that utilizes a
lock-in, a pre-amplifier, and a decade transformer to resolve the
magnetoresistance.\{(1) Decade transformer (2) $35k\Omega$ resistor
(3a) Sample resistance (3b) Contact resistances  (4) SR$560$
Pre-amplifier  (5)SR$830$ Lock-in amplifier (6) Internal lock-in
reference  (7) Lock-in inputs (8)$10$ V proportional output (9)
Oscilloscope\}}
\end{center}
\end{figure}

    Lock-in detection, nulling, four-wire measurements, and ground loops
are all essential topics for modern, condensed matter
experimentalists to be familiar with, and these labs together
provide students with a thorough, quantitative, foundation in each.
A brief schematic of our entire apparatus for performing these
functions is in Figure \ref{fig:BeyerFig7}.

\section{Typical Results}

\begin{figure}
\begin{center}
\scalebox{1.0}{\includegraphics{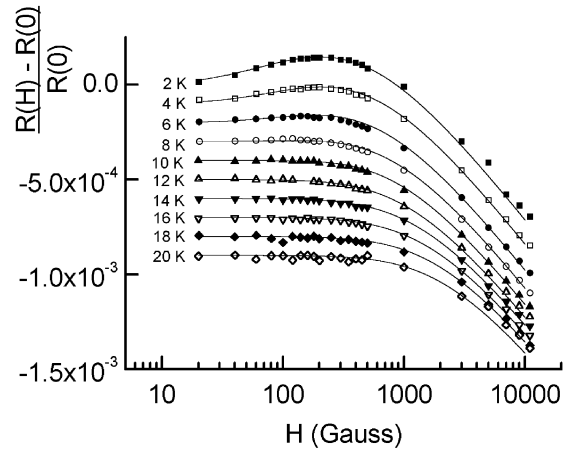}}
\caption{\label{fig:BeyerFig8} Typical data of magnetoresistance
versus field. The field scale is a log scale.  Weak localization,
with spin effects, is evident from 2K to about 14K , manifested as a
positive magnetoresistance at low fields and negative at larger
fields. Data at temperatures above 2K is offset for clarity;
magnetoresistance goes to $0$ when $H = 0$. Lines are fits to
Equation (\ref{eq:magnetoresistance}).}
\end{center}
\end{figure}

\begin{figure}
\begin{center}
\scalebox{1.0}{\includegraphics{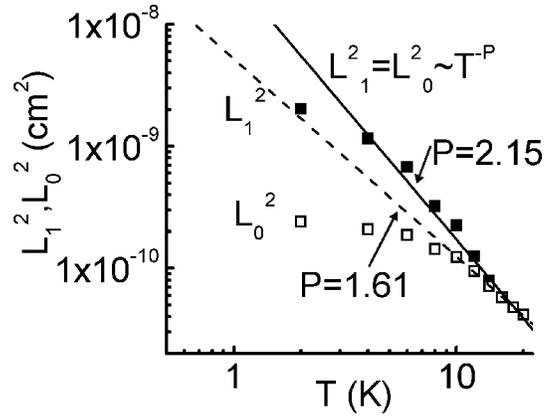}}
\caption{\label{fig:BeyerFig9} Coherence lengths squared versus
temperature.  The linear portion above 14 K has a slope between
-1.61 and -2.15. If we expect 2D thermal phonons to dominate
inelastic scattering then the slope should be P=2. }
\end{center}
\end{figure}

Using the techniques described above, our students found results
consistent with weak localization. Their results are shown in
Figures ~\ref{fig:BeyerFig8} and ~\ref{fig:BeyerFig9}.

Figure~\ref{fig:BeyerFig8} shows typical magnetoresistive data and
fits to the data.  The results clearly show weak localization with
spin effects at low temperatures (T$\leq14K$), as evidenced by the
initially positive magnetoresistance at low magnetic fields followed
by negative magnetoresistance at higher field values.  This feature
can be completely described using Equation
(\ref{eq:magnetoresistance}) to convert Equation
(\ref{eq:weak-localization}) to a magnetoresistance prediction, as
described previously, and then fitting the data. The switch from
positive to negative magnetoresistance represents a competition
between the positive magnetoresistance, due to spin effects, and the
decoherence of weak localization due to the magnetic field, which
produces a negative magnetoresistive effect. At high enough magnetic
fields, the decoherence wins out over the positive
magnetoresistance, and the sign of the magnetoresistance changes.

At T$=$14K and above, Equation (\ref{eq:weak-localization}) and
(\ref{eq:spinless-weak-localization}) can both be used to describe
the magnetoresistance (after being converted using Equation
(\ref{eq:magnetoresistance})).  This reflects the fact that spin
effects are irrelevant when phonon scattering is much more frequent
that spin-orbit scattering of electrons, which is the case at higher
temperatures (T$\geq 14K$).  AboveT$=14K$, $L_{1}\cong L_{0}$, and
if we set $L_1=L_0$ in Equation (\ref{eq:weak-localization}), then
we get Equation (\ref{eq:spinless-weak-localization}). What this
means is that we have thermally decohered the interference effects
due to spin-orbit scattering by exciting phonons.

To fit the data in Figure \ref{fig:BeyerFig8} for all temperatures,
the students performed a two parameter fit of Equation
(\ref{eq:weak-localization}) converted into magnetoresistive form by
Equation (\ref{eq:magnetoresistance}).  This required the
parameters, $L_{1}$ and $L_{0}$, to be adjusted until a good fit was
made.  We encouraged the students to fit the equations visually
until the fits were close, which was useful because it required the
students to think about the relevant length scales of $L_{1}$ and
$L_{0}$ and what they meant.  $L_{0}$ is the average length that an
electron travels before undergoing an inelastic scattering event
with a phonon. $L_{1}$, on the other hand, is a parameter that
describes whether electrons scatter more frequently from phonons or
from disorder, via spin-orbit scattering.  Their values of $L_{0}$
and $L_{1}$ for the fits in Figure \ref{fig:BeyerFig8} are shown in
Figure ~\ref{fig:BeyerFig9}, as a function of temperature.

The typical length scale for $L_1$ and $L_0$ is about $0.5\mu m$,
and at low temperatures (T$\leq14K$), $L_{1}> L_{0}$.  This
indicates that spin-orbit scattering is more frequent than phonon
scattering at low temperatures, as we expect.  This is the reason
weak localization, with spin effects, is observed.

Above 14K (as shown in Figure \ref{fig:BeyerFig9}), $L_{1}\cong
L_{0}$, and we notice that the log-log plot of $L_{0}$ and $L_{1}$
squared versus temperature has a slope between -1.61 and -2.15. In
an experimental paper with similar results to ours,
Gershenzon\cite{ref:gersh} points out that the slope of $L_1$ and
$L_0$ squared versus temperature, on a log-log plot, should be equal
to -2 if we expect 2D thermal phonons to be the source of inelastic
scattering.  So above 14K, our students are likely observing 2D
thermal phonon modes being excited and causing localization effects
to die away.

\section{Conclusions}

Our experiment on weak localization provides an introduction to
modern experimental condensed matter research. The lab illustrates
many of the experimental techniques that are used in modern research
labs. Students learn to make their own samples, make contact to
them, cool them down, and then make measurements. In addition, the
resistance bridge technique considered here is a common technique to
observe small signals on top of a large background signal.

The typical results from our students indicate that our technique is
able to observe the predicted macroscopic expression of weak
localization through the magnetoresistance of Ag thin films. Their
results also show that at easily accessible temperatures the effect
is clear to resolve.

In conclusion, the lab serves as a good transition from cookbook
type labs to the real research environment.

\section{Acknowledgments}
This work was supported by the National Science Foundation under
grant number DUE-0088658.

\bibliography{thebibliography}

\end{document}